  \documentstyle[preprint,aps]{revtex}
\begin{document}
\draft
\title{
 On the Thermodynamic Stability of Odd-Frequency Superconductors
}
\author{R. Heid}
\address{
Department of Physics,
The Ohio State University,
Columbus, Ohio 43210
}
\maketitle
\begin{abstract}
The thermodynamic stability of odd-frequency pairing states is
investigated within an Eliashberg-type framework.
We find the rigorous result that in the weak coupling
limit a continuous transition from the normal state to a spatially
homogeneous odd-in-$\omega$ superconducting state is forbidden,
irrespective of details of the pairing interaction and of the spin
symmetry of the gap function.  For isotropic systems, it is shown
that the inclusion of strong coupling corrections does not invalidate
this result.  We discuss a few scenarios that might escape these
thermodynamic constraints and permit stable odd-frequency pairing
states.

\end{abstract}


\newpage
  \narrowtext
%
%
\section{Introduction}
The puzzling physical properties of cuprate and heavy-fermion
superconductors have motivated theoretical investigations of new
classes of superconductivity which are characterized by unconventional
symmetries of the gap function $\Delta({\bf k},\omega_n)$.
A prominent example is the odd-frequency superconducting state where
$\Delta$ changes sign under inversion of frequency, originally
considered by Berizinskii in a model for triplet superfluidity in
$^3$He \cite{ber}.
This idea was picked up by Balatsky and Abrahams who proposed a new
class of odd-in-$\omega$ singlet superconductors which are odd under
parity as well due to the Pauli principle \cite{BA}.
A possible microscopic realization of odd-frequency triplet pairing has
been discussed by Coleman {\it et al.} in the context of heavy-fermion
compounds \cite{CMT}.

The physical properties of these unconventional pairing states are
rather unusual.
Because the equal-time gap vanishes,
the order parameter of the condensed state is related to the expectation
value of a composite operator \cite{BASS,Em}.
In case of a singlet pairing, the quasiparticle spectrum turns out to be
gapless \cite{BA,BASS}.
However, some concern has recently been raised with regard to the
stability of odd-frequency pairing.
As argued by Coleman {\it et al.}, a uniform s-wave triplet
odd-in-$\omega$ state should generically exhibit a negative Meissner
effect and is therefore unstable with respect to a spatial modulation
of the order parameter \cite{CMT}.
Additional support for a negative Meissner effect comes for
calculations of the Meissner kernel using normal and anomalous
Green's functions, whereas a positive Meissner effect is found within
the composite-operator description \cite{BASS}.
Furthermore, uniform odd-frequency singlet pairing seems to contradict
the requirements of causality and stability \cite{dolgov94}.

In this paper we intend to shed additional light on the question of
stability of an odd-in-$\omega$ superconductor.
This is achieved by examining the pairing induced change in the
thermodynamic potential.
The main result is that under quite general circumstances a continuous
transition from a normal state into a spatially homogeneous condensed
state is ruled out on thermodynamic grounds.
This suggests that odd-frequency pairing cannot be realized in
a uniform ground state.

The derivation of the above mentioned statement is presented in the next
section.
After introducing the formalism in Section IIA,
we first consider the weak coupling limit, where the main ideas involved
in the derivation can be presented in a very simple form (Section IIB).
However, studies of the strong-coupling gap equation show
that odd-frequency solutions often involve substantial
renormalizations in the normal channel \cite{BASS,ABSA}.
It is therefore important to include strong coupling corrections in
the stability analysis.
In Section IIC, we demonstrate that, at least for isotropic systems,
strong coupling corrections are ineffective in stabilizing a uniform
odd-frequency state.
Implications of these results are discussed in Section III.
\section{Stability analysis}
\subsection{Formalism}
We work within the framework of Fermi liquid theory.
The following assumptions are made:
 (i) no spin-orbit coupling,
 (ii) pairing occurs in a quasiparticle band described by
spin-independent energies $\epsilon_{\bf k}$ (measured from to the
chemical potential),
 (iii) the low-temperature behavior can be described by a
spin- and frequency-dependent effective interaction
$V(\sigma_1 k_1, \sigma_2 k_2 | \sigma_3 k_3, \sigma_4 k_4)$
where $\sigma_i$ are spin indices and we have used a four-vector
notation, $k=({\bf k},\omega_n$).
Fermi statistics implies that $V$ is antisymmetric with respect to the
first two and last two ($\sigma, k$) index pairs, respectively.
The only further restrictions imposed on the functional form
of $V$ come from the conservation of total spin, energy and
momentum ($\sum_{i=1}^4 k_i=0$), and from
the hermiticity of the Hamiltonian which implies
\begin{equation}
V^*(\sigma_1 k_1, \sigma_2 k_2 | \sigma_3 k_3, \sigma_4 k_4) =
V(\sigma_4 \bar{k}_4,\sigma_3 \bar{k}_3|\sigma_2 \bar{k}_2,
\sigma_1 \bar{k}_1) \label{trev}
\end{equation}
with $\bar{k}=({\bf k} , -\omega_n)$.

For a spatially homogeneous pairing state, the anomalous Green's
functions are defined as
\begin{eqnarray}
F_{\sigma \sigma^\prime}(k)=\int_0^\beta d\tau e^{i \omega_n \tau}
< T_\tau \psi_{{\bf k} \sigma}(\tau) \psi_{-{\bf k}\sigma^\prime}(0) > ,
\nonumber
\\
\hat{F}_{\sigma \sigma^\prime}(k)=\int_0^\beta d\tau e^{i \omega_n \tau}
< T_\tau \psi^\dagger_{-{\bf k} \sigma}(\tau)
\psi^\dagger_{{\bf k}\sigma^\prime}(0) > ,
\end{eqnarray}
which are related by $\hat{F}_{\sigma \sigma^\prime}(k)=
F^*_{\sigma^\prime \sigma}(\bar{k})$.
Fermi statistics requires
$F_{\sigma \sigma^\prime}(k)=-F_{\sigma^\prime \sigma}(-k)$.
Using a matrix notation, the Gor'kov equations are given by
\begin{eqnarray}
 [(i \omega_n-\epsilon_{\bf k}){\bf 1}-W(k)] G(k) &=& {\bf 1} -
 \phi(k) \hat{F}(k) ,
 \nonumber
 \\
\ [(i \omega_n+\epsilon_{\bf k}){\bf 1}+W(-k)] \hat{F}(k) &=&
  - \phi^\dagger(\bar{k}) G(k) .
\label{gorkov}
\end{eqnarray}
These equations contain the normal and anomalous parts of the electron
self-energy defined by
\begin{eqnarray}
W_{\sigma_1 \sigma_2}(k)=\frac{1}{\beta} \sum_{p \sigma_3
\sigma_4}
2 V(\sigma_1 k, \sigma_4 p | \sigma_3 p, \sigma_2 k)
G_{\sigma_3 \sigma_4}(p) \nonumber\\
\phi_{\sigma_1 \sigma_2}(k)=\frac{1}{\beta} \sum_{p \sigma_3
\sigma_4}
V(\sigma_1 k, \sigma_2 -k | \sigma_3 p, \sigma_4 -p)
F_{\sigma_3 \sigma_4}(p) .
\label{gap}
\end{eqnarray}
As a consequence of the antisymmetry of $V$,
the relation
$\phi_{\sigma\sigma^\prime}(k)=-\phi_{\sigma^\prime\sigma}(-k)$
holds.

The following stability analysis of the odd-frequency
superconducting state makes use of a general expression for the change
in the thermodynamic potential due to the two-body interaction $V$
\cite{AGD}
\begin{equation}
\delta\Omega = \frac{1}{\beta}\int_0^1 \frac{dg}{2g} \sum_k tr \{(i
\omega_n-\epsilon_{\bf k}) G(k) - {\bf 1}\}_g .
\label{deff}
\end{equation}
The trace is taken over spin indices, and the index $g$ indicates that
the quantities in the bracket correspond to
a system with a scaled interaction $gV$.
This formula is not restricted to weak coupling and
is applicable even in the case of a critical lower coupling strength.
We assume that the interaction potential $V$ allows a certain type of
solution of the gap equations (\ref{gorkov}) and (\ref{gap}), and
examine its thermodynamic stability with respect to the normal state.
It is now useful to distinguish between weak and strong coupling.


\subsection{Weak coupling}
The weak coupling approximation is obtained by neglecting the
self-energy contributions in the normal channel, $W=0$,
in which case $\delta\Omega$ equals the difference between the
normal and superconducting state.
Combining (\ref{gorkov}) and (\ref{deff}), we obtain
\begin{mathletters}
\begin{eqnarray}
\delta\Omega &=& -\frac{1}{\beta}\int_0^1 \frac{dg}{2g} \sum_k
tr \{\phi(k)\hat{F}(k)\}_g
\label{doma}
\\
&=& -\frac{1}{\beta}\int_0^1 \frac{dg}{2g} \sum_k tr \left\{
[( \omega_n^2+\epsilon_{\bf k}^2){\bf 1} +
\phi^\dagger(\bar{k}) \phi(k)]^{-1}
\phi^\dagger(\bar{k}) \phi(k)
\right\}_g .
\label{dom}
\end{eqnarray}
\end{mathletters}
Let us now consider a continuous phase transition. For temperatures
close to $T_c$ we can neglect the term proportional
to $\phi^2$ in the denominator
  \begin{equation}
\delta\Omega =-\frac{1}{\beta}\int_0^1 \frac{dg}{2g} \sum_{{\bf
k},\omega_n} \frac{
tr\{\phi^\dagger({\bf k},-\omega_n) \phi({\bf k},\omega_n)\}_g}
{\omega_n^2 +\epsilon_{\bf k}^2} .
\label{omweak}
\end{equation}
In this form the sign of $\delta\Omega$ can be determined by solely
relying on symmetry properties of the gap function.
There are two distinct cases:
\begin{enumerate}
 \item $\phi({\bf k},-\omega_n)=\phi({\bf k},\omega_n)$,
i.e. $\phi$ is even in $\omega$.
Because $\phi^\dagger({\bf k},\omega_n) \phi({\bf k},\omega_n)$ is a
positive definite matrix (for all $g$), it follows that
$\delta\Omega<0$ and the paired state is stable with respect to the
normal state.
 \item $\phi({\bf k},-\omega_n)=-\phi({\bf k},\omega_n)$,
i.e. $\phi$ is odd in $\omega$.
This implies $\delta\Omega>0$ showing that in this case the condensed
state does not fulfill the thermodynamic stability criterion.
\end{enumerate}
Consequently, a
second-order transition to a spatially homogeneous, odd-frequency
superconducting state is forbidden.
Furthermore, in the weak coupling limit, this
conclusion does not depend on the details of the pairing interaction
and is valid for singlet and triplet pairing.


\subsection{Strong coupling}
We now extend the above given arguments to incorporate strong coupling
corrections.
In this case, no derivation with a generality similar to the weak
coupling limit exists. However, we show that the same conclusions still
hold for the class of isotropic models.
This class is defined by
\begin{itemize}
\item[(i)] isotropic quasiparticle energies, $\epsilon_{\bf
k}=\epsilon_{|\bf k|}$, and particle-hole symmetry;
\item[(ii)] a general spin-dependent interaction
$V(12|34)=\Gamma(1234)-\Gamma(1243)$ with
\begin{equation}
\Gamma(1234)=\left[ \Gamma^C(k_1-k_3) \delta_{\sigma_1\sigma_3}
\delta_{\sigma_2\sigma_4}
+\Gamma^S(k_1-k_3) \sum_i \sigma^i_{\sigma_1 \sigma_3} \sigma^i_
{\sigma_2 \sigma_4} \right]
\delta_{k_1-k_3,k_4-k_2}
\end{equation}
where $1\equiv (k_1\sigma_1)$, etc.\ , and $\sigma^i$, $i=x,y,z$, denote
Pauli matrices.
$\Gamma^C$ and $\Gamma^S$ are functions symmetric in frequency and
rotationally invariant in momentum space;
\item[(iii)] negligible $|{\bf k}|$-dependence of $\Gamma$ and the
self-energy in the vicinity of the Fermi surface.
\end{itemize}
The same type of interaction has been used in the work of
Abrahams {\em et al.} \cite{ABSA}.

For singlet pairing the normal
self-energy is diagonal in spin-space, $W(k)=i\omega_n Z(k){\bf 1}$,
and the gap function $\Delta$ is given by $\phi(k)=i\sigma^y Z(k)
\Delta(k)$.
Under these conditions, (\ref{deff}) is equivalent to
\begin{equation}
\delta\Omega=-\int_0^1\frac{dg}{g}
\left\{
{\cal N}(0)\frac{\pi}{\beta}\sum_n
\int\frac{d\Omega_{\hat{\bf k}}}{4\pi}
\frac{\omega_n^2(Z(\hat{\bf k},\omega_n)-1)+Z(\hat{\bf k},\omega_n)
\Delta(\hat{\bf k},\omega_n)\Delta^*(\hat{\bf k},-\omega_n)}
{\sqrt{\omega_n^2+\Delta(\hat{\bf k},\omega_n)\Delta^*
(\hat{\bf k},-\omega_n)}} \right\}_g ,
\label{str1}
\end{equation}
where ${\cal N}(0)$ is the density-of-states per spin at the Fermi
energy, and $\hat{\bf k}$ is a vector on the Fermi surface.
The renormalization factor takes the form
\begin{equation} Z(\hat{\bf k},\omega_n)=1+\frac{1}{\omega_n}
\frac{\pi}{\beta}\sum_m \int \frac{d\Omega_{\hat{\bf p}}}{4 \pi}
\gamma(\hat{\bf k}\cdot\hat{\bf p},\omega_n-\omega_m) \frac{\omega_m
}{\sqrt{\omega_m^2+\Delta(\hat{\bf p},\omega_m)
\Delta^*(\hat{\bf p},-\omega_m)}}
\label{zeq}
\end{equation}
Here, $\gamma=2{\cal N}(0)(\Gamma^C+3\Gamma^S)$ is the effective
interaction entering the normal self-energy.

The possibility of a second order transition from the normal state into
the superconducting state is determined by the sign of the second order
term in an expansion of $\delta\Omega$ in the gap function $\Delta$.
Using (\ref{zeq}) to replace $Z$, one finds
\begin{eqnarray}
\delta\Omega_2
&=&-\int_0^1\frac{dg}{g}
\left\{
{\cal N}(0)\frac{\pi}{\beta}\sum_n
\int\frac{d\Omega_{\hat{\bf k}}}{4\pi}
\frac{\Delta(\hat{\bf k},\omega_n)\Delta^*(\hat{\bf k},-\omega_n)}
{|\omega_n|} \right. \nonumber\\
&&+\frac{\pi^2}{\beta^2}\sum_{n m}
\left.
\int\frac{d\Omega_{\hat{\bf k}}}{4\pi}\int\frac{d\Omega_{\hat{\bf p}}}
{4\pi} \frac{\omega_n}{|\omega_n|}\gamma(\hat{\bf k}\cdot\hat{\bf p},
\omega_n-\omega_m) \frac{\omega_m}{|\omega_m|}\right. \nonumber\\
&&\hspace{.4in}\times\left.\left(\frac{\Delta(\hat{\bf k},\omega_n)
\Delta^*(\hat{\bf k},-\omega_n)}
{\omega_n^2}-\frac{\Delta(\hat{\bf p},\omega_m)
\Delta^*(\hat{\bf p},-\omega_m)}{\omega_m^2}\right)
\right\}_g
\end{eqnarray}
Because $\gamma$ is symmetric with respect to an exchange
$(\hat{\bf k}\omega_n) \leftrightarrow (\hat{\bf p}\omega_m)$, the last
term vanishes when summed over frequencies, and we are left with
\begin{equation}
\delta\Omega_2
=-\int_0^1\frac{dg}{g}
\left\{
{\cal N}(0)\frac{\pi}{\beta}\sum_n
\int\frac{d\Omega_{\hat{\bf k}}}{4\pi}
\frac{\Delta(\hat{\bf k},\omega_n)\Delta^*(\hat{\bf k},-\omega_n)}
{|\omega_n|} \right\}_g \\
\label{str4}
\end{equation}
The expression for $\delta\Omega_2$ is very similar to
(\ref{omweak}), and application of the same reasoning as in the
previous subsection shows that strong coupling corrections cannot
stabilize odd-in-$\omega$ pairing for the class of models considered.
This result is in agreement with the conclusions obtained by Dolgov and
Losyakov \cite{dolgov94}.

The present derivation is not restricted to singlet pairing, but
holds for unitary triplet pairing, too, where the normal self-energy
retains the same spin-diagonal form, and $\phi(k)=iZ(k)({\bf
d}(k)\cdot\vec{\sigma})\sigma^y$ with ${\bf d}(k)\times {\bf
d}^*(\bar{k})=0$.  In this case, the product
$\Delta(k)\Delta^*(\bar{k})$ is simply replaced by the scalar product
${\bf d}(k)\cdot{\bf d}^*(\bar{k})$ in
(\ref{str1})-(\ref{str4}).  No similar derivation seems to exist,
however, for a nonunitary triplet pairing (${\bf d}(k)\times {\bf
d}^*(\bar{k})\neq 0$).

\section{Discussion}

The result obtained in the previous section emphasizes some constraints
imposed by thermodynamics which must be fulfilled by any realization of
these unconventional superconducting states.
However, our derivation should not be mistaken as a proof of the
nonexistence of odd-frequency pairing in general.
As mentioned before, it requires the applicability of an
Eliashberg-type approach.
Moreover, there are certain scenarios which are not covered by our
analysis, and may allow stable odd-in-$\omega$ states.

First, it is possible that in a system with large anisotropy,
strong coupling corrections are more effective in producing a stable
uniform odd-frequency superconducting ground state.
Second, our analysis concentrated on second order transitions
only, but does not exclude a first order transition.
This possibility is
easiest seen for singlet pairing in the weak coupling limit (i.e.
$\phi(k)=i\sigma^y\Delta(k)$).
In the odd-in-$\omega$ case, the $\Delta^2$ term in the
denominator of (\ref{dom}) is negative, leading to
\begin{equation}
\delta\Omega =
\frac{1}{\beta}\int_0^1 \frac{dg}{g} \sum_k \left\{
\frac{|\Delta(k)|^2}{ \omega_n^2+\epsilon_{\bf k}^2 -
 |\Delta(k)|^2}
 \right\}_g .
\end{equation}
If the sign in the denominator is changed for a sufficiently large
$k$-region, the pairing state might be stabilized.

A third scenario is based on the idea of a spatial
inhomogeneity of the pairing amplitude.
Here, the Cooper pairs acquire a finite center-of-mass momentum ${\bf
q}$,
which describes the modulation of the order parameter on the lattice.
This type of pairing has recently been discussed in more detail in the
context of heavy fermion superconductors, and is thought to be driven
by the negative Meissner stiffness of the homogeneous odd-in-$\omega$
state, which favors a coiling of the order parameter phase
\cite{CMT,heid}.

It is easy to connect this picture to the thermodynamic
stability consideration.
Let $\delta\Omega({\bf q})$ be the potential
difference between the normal state and a superconducting state with a
finite total momentum ${\bf q}$ of the electron pairs.
The Meissner stiffness is proportional to the coefficients of the
gradient terms in a Ginzburg-Landau expansion of the free energy, and
is therefore related to the second derivative of $\delta\Omega({\bf
q})$ with respect to ${\bf q}$.
The positive Meissner stiffness found for a
homogeneous phase of an even gap superconductor
thus corresponds to a minimum of
$\delta\Omega({\bf q})$ at ${\bf q}=0$.
The results of the previous section, in conjunction with a negative
Meissner stiffness obtained for the uniform odd-in-$\omega$ state,
suggests that in this case $\delta\Omega({\bf q})$ is instead maximal
at ${\bf q}=0$.
Consequently, the global minimum of $\delta\Omega({\bf q})$ occurs
at a finite momentum ${\bf q}_{min}$, which defines an odd
superconducting state with a positive Meissner stiffness.
This state is stable with respect to the normal state if
$\delta\Omega({\bf q}_{min})<0$, a necessary condition which may be
violated even for interactions attractive in the odd-frequency pairing
channel (i.e. which allows odd-in-$\omega$ solutions of the gap
equations).
This contrasts the even-in-$\omega$ case, where a homogeneous phase
is always stable for an attractive pairing potential.

One should keep in mind that the given discussion solely addresses the
thermodynamic stability of odd-in-$\omega$ states with respect to the
normal state.
States which are stable according to this criterion do not necessarily
correspond to the ground state, because competing even-frequency states
might possess even lower free energies.
This question can only be answered when a specific interaction potential
is given.

In summary, we have shown that the thermodynamic stability criterion
puts severe restrictions on the occurrence of a continuous phase
transition to a spatially homogeneous odd-in-$\omega$ superconducting
state.
This result strongly suggests that any realistic model of this
unconventional pairing requires a spatial modulation of the
order parameter on an atomic scale from the very beginning.

\acknowledgements
I gratefully acknowledge stimulating discussions with Daniel Cox and
Michael Reyzer.
This work was supported by
the Deutsche Forschungsgemeinschaft.

\end{document}